\documentclass[aps,pra,showpacs,twocolumn,superscriptaddress]{revtex4-1}

\usepackage{graphicx,color}
\usepackage{amsmath,amsthm,amsfonts,amssymb,bm}
\usepackage{times}
\usepackage[colorlinks={true}]{hyperref}

\usepackage[T1]{fontenc}
\usepackage[utf8]{inputenc}
\usepackage{ulem}

\hypersetup{citecolor={blue}, filecolor={blue}, linkcolor={blue}, urlcolor={blue}}

\usepackage{graphicx}
\begin{document}

\title{Non-Markovianity, coherence and system-environment correlations in a long-range collision model}

\author{B. \c{C}akmak}
\email{bcakmak@ku.edu.tr}
\affiliation{Department of Physics, Ko\c{c} University, \.{I}stanbul, Sar\i yer 34450, Turkey}
\author{M. Pezzutto}
\email{marco.pezzutto@tecnico.ulisboa.pt}
\affiliation{Instituto de Telecomunicações, Physics of Information and Quantum Technologies Group, Lisbon, Portugal}
\affiliation{Instituto Superior Técnico, Universidade de Lisboa, Portugal}
\affiliation{Centre for Theoretical Atomic, Molecular and Optical Physics, School of Mathematics and Physics, Queen’s University, Belfast BT7 1NN, United Kingdom}
\author{M. Paternostro}
\affiliation{Centre for Theoretical Atomic, Molecular and Optical Physics, School of Mathematics and Physics, Queen’s University, Belfast BT7 1NN, United Kingdom}
\author{\"{O}. E. M\"{u}stecapl{\i}o\u{g}lu}
\affiliation{Department of Physics, Ko\c{c} University, \.{I}stanbul, Sar\i yer 34450, Turkey}

\begin{abstract}
We consider the dynamics of a collisional model in which both the system and environment are embodied by spin-$1/2$ particles. In order to include non-Markovian features in our model we  introduce interactions among the environmental qubits and investigate the effect that different models of such interaction have on the degree of non-Markovianity of the system's dynamics. By extending that interaction beyond the nearest-neighbour, we enhance the degree of non-Markovianity in the system's dynamics. A further significant increase can be observed if a collective interaction with the forthcoming environmental qubits is considered. However, the observed degree of non-Markovianity in this case is non-monotonic with the increasing number of qubits included in the interaction. Moreover, one can establish a connection between the degree of non-Markovianity in the evolution of the system and the fading behaviour of quantum coherence in its state as the number of collisions grow. We complement our study with an investigation on system-environment correlations and present an example of their importance on a physical upper bound on the trace distance derivative. 
\end{abstract}


\date{\today}

\maketitle


The theory of open quantum systems deals with the inevitable interaction between a system and its surrounding environment, which results in a non-unitary time evolution of the system density matrix~\cite{bbook,book2,book3}. As a result, quantum coherence and the information encoded in the system's state are lost into the environmental degrees of freedom. In the case of a Markovian evolution, the loss of system information is monotonic and at any time the future evolution of the system only depends on its present state. On the other hand, non-Markovian dynamics can be associated with a temporary reverse of such a flow of information, which results in the system regaining some of the lost information and making the future evolution of the system dependent on its past.

Recently, the characterization and quantification of non-Markovian dynamics has attracted a lot of attention. The tools of quantum information theory have been used extensively to quantify the amount of information backflow from the environment to the system, thus providing an important intuitive understanding of non-Markovianity \cite{blp,measures}. The {\it measures of non-Markovianity} put forward so far are helping us characterize the features of memory-bearing quantum open-system dynamics, shedding light on the ultimate origins of such behaviours. 
In general, however, they do not mutually agree on the emergence and degree of non-Markovianity. In this sense, they all characterize it from different perspectives \cite{compare}.

In this work, we consider a  model that describes the system-environment interaction through a series of sequential "collisions" between the system and the environmental particles. Such "collisional" model is capable of simulating both Markovian and non-Markovian dynamics depending on the interaction and/or correlation between the environmental degrees of freedom \cite{homogenization, ziman,rybar,mccloskey,ciccarello,ciccarello2,kretschmer}. We explore how various ways of engineering the interactions among the particles in the environment affect the degree of non-Markovianity of the dynamics of the system. In particular, we consider separate and collective long-range interactions and determine if and how the non-Markovianity, as quantified using the tool put forward in Ref.~\cite{blp}, is affected by the different ways in which the information propagates through the environment. Furthermore, we investigate the relation between the amount of non-Markovianity that our dynamical model generates and the behavior of the coherence in the system's state. Lastly, we turn our attention to the relation between system-environment correlations and non-Markovianity, which are believed to be intimately related.

The analysis reported in this work allows us to highlight a set of counterintuitive results. First, we find that the degree of non-Markovianity of the dynamics of the system appears to be decreasing with the depth of the collective interactions considered in our study. In fact, we show how the inclusion of  non-nearest neighbour interactions does not necessarily result in a more pronounced non-Markovian character of the dynamics, as one might expect. Moreover, we unveil a peculiar relation between quantum coherence and the {\it nature} of the coupling with the environment: while we find vanishing quantum coherences for single-environment interactions, the coupling to a collective environment appears to shield them for more than an order of magnitude higher number of collisions. Such a protection effect shows direct proportionality with the degree of non-Markovianity of the dynamics. 

The remainder of the paper is organized as follows. Sec.~\ref{model} introduces the general idea behind collisional models and gives the specifications of the models we consider throughout this manuscript. We also briefly introduce the measure that is going to be utilized to quantify non-Markovianity of the dynamics in the same section. Sec.~\ref{results} presents the results on the non-Markovian features of the dynamics produced by our collision model and its effect on coherence and system-environment correlations. Finally, in Sec.~\ref{conclusions} we draw our conclusions.

\section{Collision model}
\label{model}

The collisional model that we consider consists of a system ($s$) interacting with an ensemble of environmental particles, $e\in\{e_1, e_2, \cdots, e_m\}$, one at a time. Each $e_k$, $k\in\{1,\cdots, m\}$, is a {\it subenvironment}, and $m$ is the number of elements constituting the environment. Throughout this work we will take each $e_k$ to be a two-level system. In our model, a given subenvironment interacts with the system only once and  is then discarded.


In any one step of the dynamics, the system interacts with the $k^{\text{th}}$ subenvironment, which is then coupled to the  forthcoming subenvironment. In order to obtain the reduced state of the system we trace out the environment. Repeating this process in an iterative loop for the desired number of times results in the full time evolution of the system qubit. The dynamical maps that governs this time evolution can be written as
\begin{equation}
  \Lambda[\rho] = U_{se}\rho U_{se}^{\dagger},~~~ 
  \Psi[\rho] = U_{ee}\rho U_{ee}^{\dagger}, \label{eq:Uee}
\end{equation}
where $U_{se}=\exp(-iH_{se}\phi)$ and $U_{ee}=\exp(-iH_{ee}\theta)$ with $H_{se}$ ($\phi$) and $H_{ee}$ ($\theta$) the  $s$-$e$ and $e$-$e$ interaction Hamiltonians (strengths), respectively. If we initialize the collective system and environmental state in the factorized form $\rho^{se}_0=\rho^s_0\otimes\rho^e_0$, it is possible to obtain the final combined state after the $k^{\text{th}}$ iteration by a unitary transformation
\begin{equation}\label{rhokth}
  \rho^{se}_{k}=U\rho^{se}_{0} U^{\dagger},
\end{equation}
where $U$ is composed of sequential applications of $U_{se}$ and $U_{ee}$. The reduced state of the system can be obtained by tracing out the environmental degrees of freedom, $\rho^s_{k}={\rm Tr}_e(\rho^{se}_{k})$.

In the case of identical non-interacting  subenvironments, one gets a dynamical process called quantum homogenization \cite{homogenization}. As the system qubit collides with the environmental ones, its state will gradually change, and after a sufficient number of collisions it will eventually become identical to the initial state of the subenvironments. Clearly, this is a microscopic model of Markovian decoherence. However, adding $e$-$e$ collisions to the model, the reduced dynamical evolution of $s$ becomes non-Markovian: some of the information that the system has lost to a subenvironment propagates within the environment due to $e$-$e$ interactions, and is fed back to the system at a later collision. The form of $H_{ee}$ strongly affects the degree of non-Markovianity of the dynamics. For example, a SWAP-like interaction between neighbouring subenvironments results in a non-Markovian time evolution whose degree is determined by whether a partial or full SWAP operation is used \cite{mccloskey}.
\begin{figure}[h]
  \centering
  \includegraphics[width=.48\textwidth]{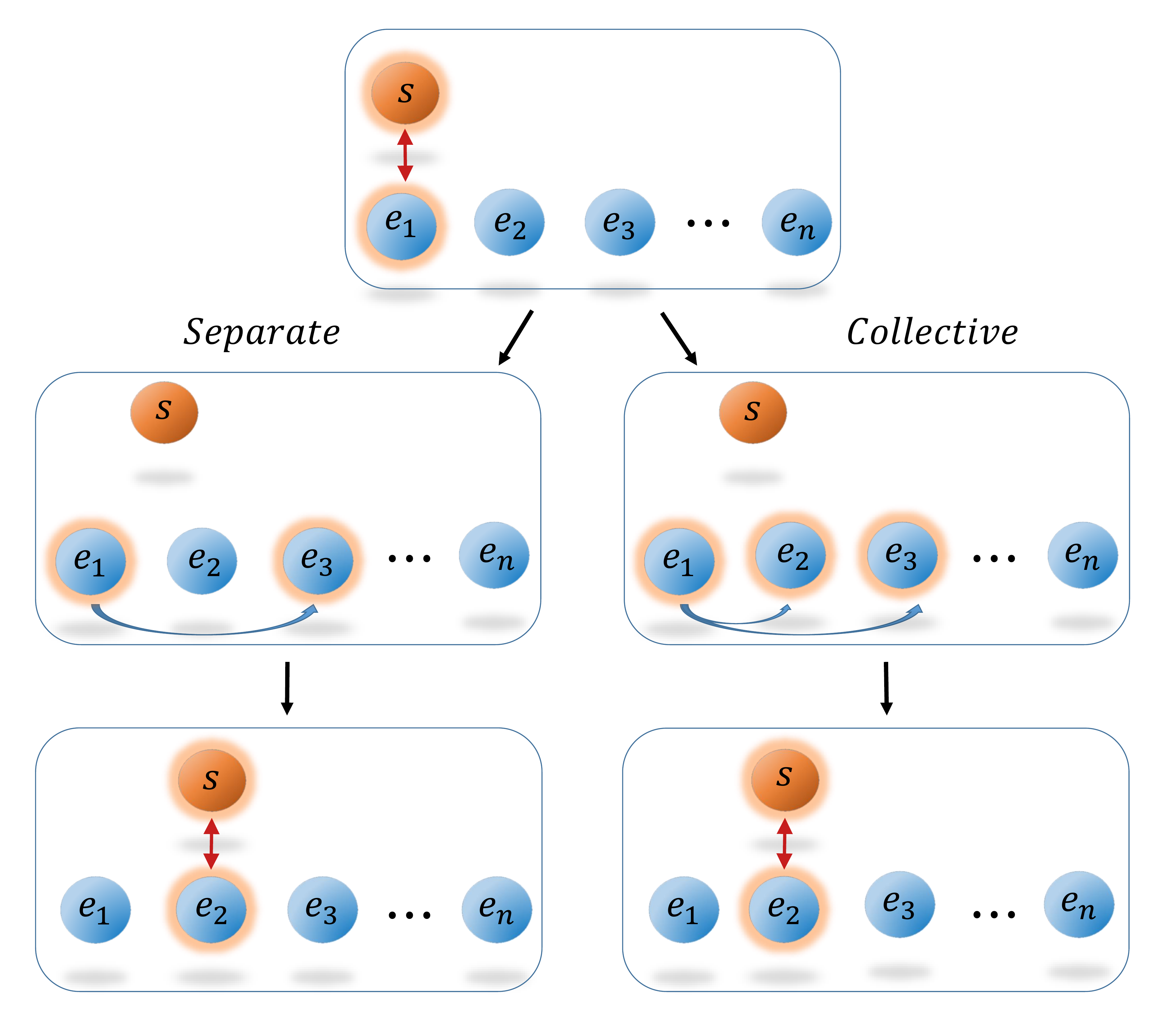}
  \caption{Schematic view of the two different environmental interaction models considered in this work. The left column describes the separate interaction with the $2^{nd}$ nearest-neighbour, while the right column depicts the collective interaction scenario with qubits upto $2^{nd}$ nearest-neighbour. Generalization to longer-range interactions follows from the picture presented here. \label{fig:model}}
\end{figure}

In this work, our aim is to construct a simple collisional model which allows for demonstration and control of non-Markovian features. We model the $s$-$e$ and $e$-$e$ couplings as spin-spin interactions, which may be implemented in systems like quantum dot spin-valve type devices (see e.g. \cite{strasberg} and references therein) or molecular nano-magnets. In order to model the $s$-$e$ interaction, we choose the Hamiltonian governing the dynamics as (we take units such that $\hbar=1$ throughout the manuscript)
\begin{equation}\label{hse}
  H_{se}=J_{se}(\sigma_x^s\sigma_x^e+\sigma_y^s\sigma_y^e),
\end{equation}
where $\sigma_{x,y,z}$ are the Pauli matrices. 
For the case of $e$-$e$ coupling, we want to extend the length of the interaction beyond nearest-neighbour ($nn$) and see if and how the amount of non-Markovianty depends on such a modification. We introduce the long-range interactions in two ways, as depicted in Fig.~\ref{fig:model}. On one hand, we consider separate interactions with $2^{\text{nd}}$, $3^{\text{rd}}$ or $4^{\text{th}}$ $nn$ subenvironments, for which the $e$-$e$ Hamiltonians can be written as the Heisenberg-like couplings
\begin{equation}
  H_{ee}^j = {J_{ee}}\sum^{m-j}_{i=1}(\sigma_x^{e_i}\sigma_x^{e_{i+j}}+\sigma_y^{e_i}\sigma_y^{e_{i+j}}+\sigma_z^{e_i}\sigma_z^{e_{i+j}})/2
\end{equation}
with $j=1,..,4$. On the other hand, we introduce subenvironment interactions as an equally weighted linear combination of the $H_{ee}^i$, $i=1, 2, 3, 4$, such as $H_{ee}^{12}=H_{ee}^1+H_{ee}^2$, $H_{ee}^{123}=H_{ee}^1+H_{ee}^2+H_{ee}^3$ and $H_{ee}^{1234}=H_{ee}^1+H_{ee}^2+H_{ee}^3+H_{ee}^4$. This scenario can be seen as the collective interaction of the environmental qubit which has interacted with the system, with the remaining environmental qubits. Interactions between subenvironments are designed to be only in the forward direction, i.e. $e_i$ interacts with $e_i'$ particle(s) only if $i'>i$ as the environmental particles with $i'<i$ have already been discarded. Throughout this work, we solve the dynamics for system particles numerically since, analytical approaches, such as deriving a master equation for the cases addressed here, is far from tractable.

In order to quantify and discuss the non-Markovian behaviour in our models, it is now appropriate to introduce the measure of non-Markovianity which will be utilized in this manuscript. It is known as the BLP measure \cite{blp} and based on the trace distance between two quantum states
\begin{equation}\label{td}
  D(\rho_1(t),\rho_2(t))=\frac{1}{2}||\rho_1(t)-\rho_2(t)||_1,
\end{equation}
where $||.||_1$ is the trace norm. The trace distance is zero for indistinguishable (identical) quantum states, while it is unity for completely distinguishable (orthogonal) quantum states, thus it can be thought as measure of distinguishability.

Consider two completely distinguishable initial states which are then exposed to same Markovian environment. Both of the initial states will eventually lose all their initial information into the environmental degrees of freedom, and become identical. Due to the Markovian nature of the dynamical process, the loss of their distinguishability, as quantified by the trace distance, will be monotonic in time. In other words, the rate of change of the trace distance will always be negative, $dD/dt<0$. However, if there is a deviation from this behavior, such that $dD/dt>0$, we can conclude that the dynamical evolution under consideration is non-Markovian in nature. Intuitively, we can interpret the increase in the trace distance as a backflow of the information that the subject system has lost into the environment. Based on this, it is possible to define a measure of non-Markovianity as follows \cite{blp}
\begin{equation}\label{blp}
  \mathcal{N}=\max_{\rho_1(0), \rho_2(0)}\int_{dD/dt>0}\frac{dD}{dt}dt,
\end{equation}
where the maximization is made over all possible pairs of initial states $\rho_1(0)$ and $\rho_2(0)$. Since in the collision model considered in this work the time evolution takes place in discrete steps, we will use the discretized version of the above measure,  expressed as \cite{laine,vacchini}
\begin{equation}\label{dblp}
  \mathcal{N}=\max_{\rho^s_{1,0}, \rho^s_{2,0}}\sum_{k}[D(\rho^s_{1,k},\rho^s_{2,k})-D(\rho^s_{1,k-1},\rho^s_{2,k-1})],
\end{equation}
where $k$ is the index that denotes the collision number. It is important to note that, while observing a temporary increase in the trace distance is sufficient to conclude that the dynamical map is non-Markovian, the converse statement is not always true: there may be a non-Markovian time evolution in which the trace distance decreases monotonically. In this sense, the condition $dD/dt>0$ is only a witness for non-Markovianity.

\section{Results}
\label{results}
\subsection{Non-Markovian evolution}

\subsubsection{Separate interaction case}

We start by presenting our findings on separate $2^{\text{nd}}$, $3^{\text{rd}}$ and $4^{\text{th}}$ $nn$ subenvironment interactions and compare them to the case of $nn$ coupling. To begin with, assume weak system-environment coupling ($J_{se}t\ll{1}$) and set the interaction strength $J_{se}t=0.05$, where $t$ is the interaction time. Considering the interactions between the subenvironments, it is known that for the $nn$ interaction the maximum degree of non-Markovianity is obtained when $J_{ee}t=\pi/2$, which upto a global phase, corresponds to the full SWAP operation between the neighbour environmental systems~ \cite{mccloskey}. This is a rather expected result, as by completely swapping the two subenvironments one actually makes the system interact with the same environmental state at every $s$-$e$ collision. Decreasing the value of $J_{ee}t$ below $\pi/2$ will result in the gradual degradation and eventual loss of the non-Markovian features of the model. The same line of thought also applies in the case of distant $e$-$e$ couplings: to obtain the highest degree of non-Markovianity we again set $J_{ee}t=\pi/2$. The initial system states that maximize the BLP measure are $|\pm\rangle=(|0\rangle\pm|1\rangle)/\sqrt2$ and the initial state of each subenvironment is set to be $|0\rangle$.

\begin{figure}[h]
  \centering
  \includegraphics[width=.48\textwidth]{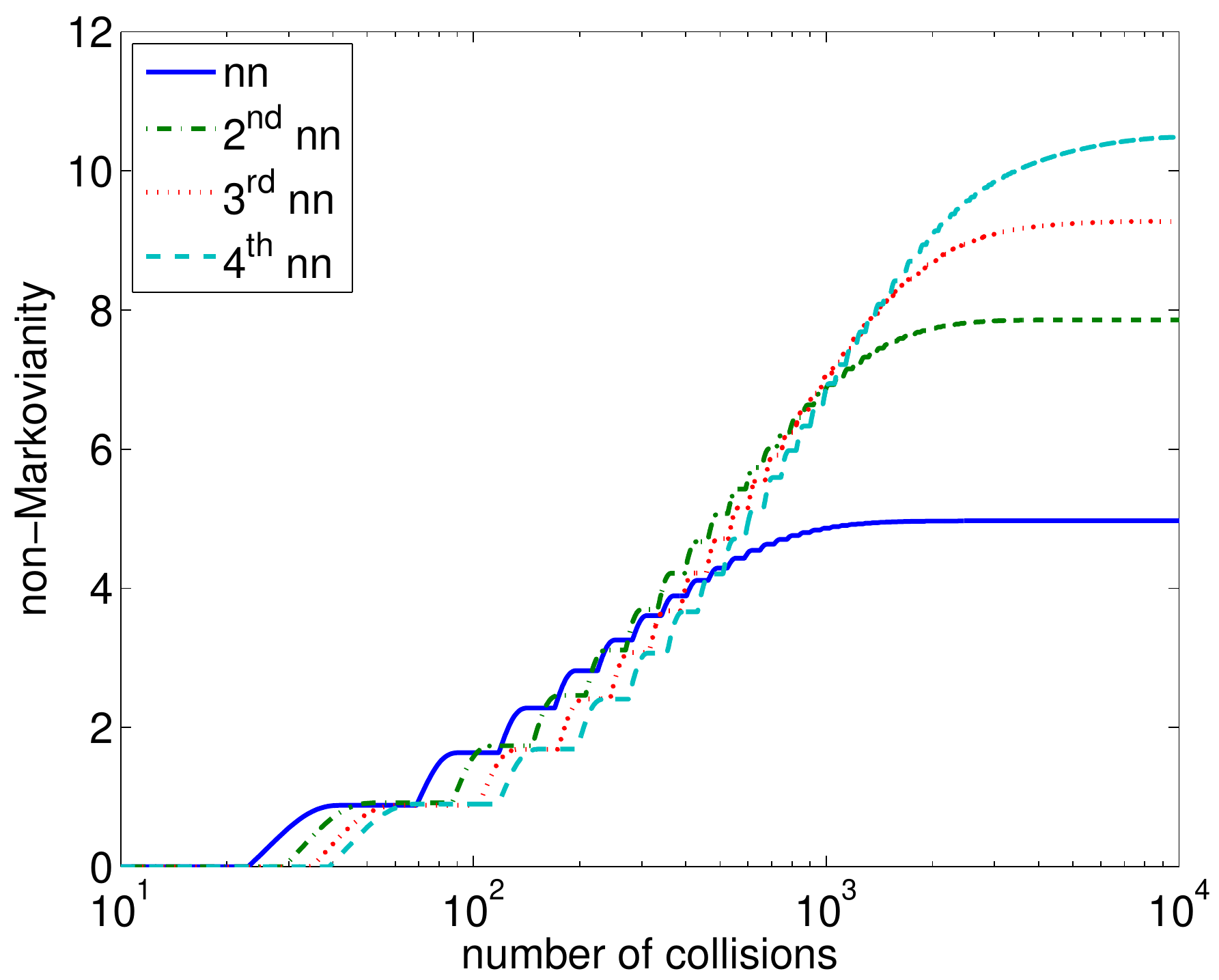}
  \caption{Non-Markovianity against the number of collisions for $nn$, $2^{\text{nd}}$ $nn$, $3^{\text{rd}}$ $nn$, $4^{\text{th}}$ $nn$ interactions. The initial system states that maximize $\mathcal{N}$ are $|\pm\rangle$ and $J_{ee}t=\pi/2$ for all cases. \label{fig:separate}}
\end{figure}

It can be seen in Fig.~\ref{fig:separate} that, as the distance between the two interacting subenvironments increases  up to the $4^{\text{th}}$ $nn$, the degree of non-Markovianity monotonically increases too. Furthermore, we observe that the number of collisions needed for the saturation of $\mathcal{N}$ is affected by the choice of the environmental interaction: the system qubit needs to go through a higher number of collisions as compared to the $nn$ $e$-$e$ interaction case, before settling to a final state which is the same as the environment initial state.

It is also important to note that, as the distance between two interacting subenvironments increases, we observe a shift towards higher values in the number of collisions needed, to have a non-zero degree of non-Markovianity. The reason behind this is that in the cases of $2^{\text{nd}}$, $3^{\text{rd}}$ and $4^{\text{th}}$ $nn$ interactions, the system qubit has to interact respectively with one, two and three subenvironments, which are all in their initial state, before it comes in contact with the subenvironment that has a partial information about its past state.

As an extension to the single separate interaction, it is possible to consider two or more consecutive separate interactions of the environmental qubits with different or same coupling strengths. However, these scenarios do not increase the non-Markovianity due to the SWAP-like form of the interaction that we choose in our model. For example, considering a $nn$ interaction followed by a $2^{\text{nd}}$ $nn$ interaction with $J_{ee}t=\pi/2$ will result in the same value of $\mathcal{N}$ as the sole $nn$ interaction: when the interaction with the $2^{\text{nd}}$ $nn$ takes place, the environmental qubit that has interacted with the system has already been swapped with the $nn$ subenvironment, which is in its initial state. Therefore the $2^{\text{nd}}$ $nn$ interaction does not produce any difference in the dynamics. One can naturally ask what happens if we relax the full SWAP condition and freely change the interaction strengths of various consecutive separate interactions. Even though the answer is not quite definitive, one can still observe, following the example above, that the degree of non-Markovianity changes from zero to the maximum obtained in the sole $2^{\text{nd}}$ $nn$ interaction. Therefore, we can tune the amount of non-Markovianity of the dynamical evolution in our model. This example can be generalized to any combination of consecutive interactions considered in this work, given that the interactions are ordered by the increasing distance between environmental qubits.

\subsubsection{Collective interaction case}

We would like now to turn our focus on the collective long-range interaction model, and again see whether it is possible to enhance the degree of non-Markovianity, as quantified by the trace distance. We again assume weak system-environment coupling $J_{se}t=0.05$. Comparing with the previous case, we see that the value of $J_{ee}t=\pi/2$ which yields the strongest non-Markovianity for the separate interaction scenario, is no longer the best choice if we change the environmental interaction Hamiltonian to a collective one. In fact, we find that the maximum degree of non-Markovianity is obtained for $J_{ee}t=0.6(\pi/2)$, $J_{ee}t=0.43(\pi/2)$ and $J_{ee}t=0.33(\pi/2)$ for $H_{ee}^{12}$, $H_{ee}^{123}$ and $H_{ee}^{1234}$, respectively. Moreover, apart from these specific $J_{ee}t$ values, $\mathcal{N}$ is always zero, which implies that either the time evolution is Markovian, or that its non-Markovian character can not be detected by the BLP measure.

\begin{figure}[t]
  \centering
  \includegraphics[width=.48\textwidth]{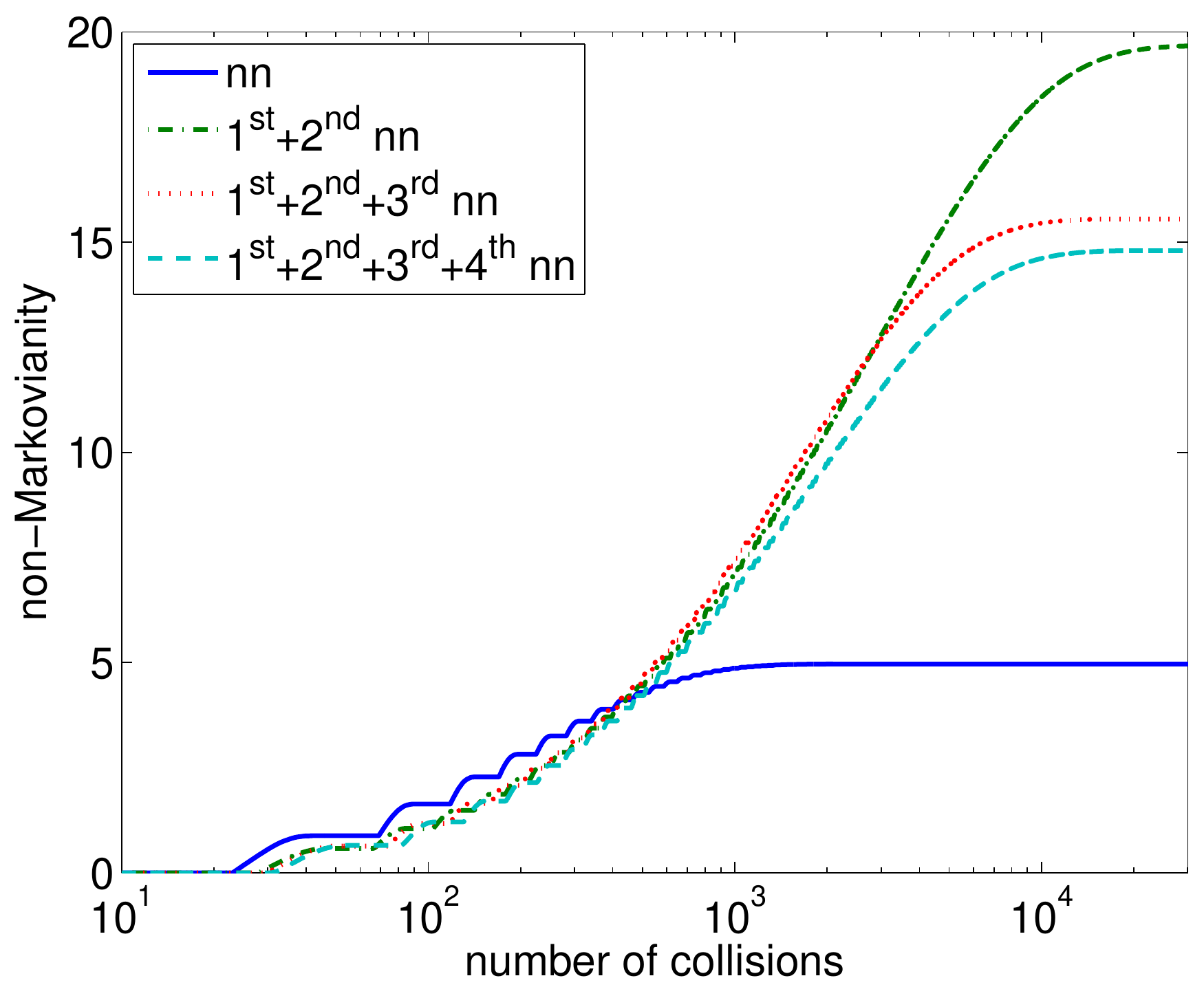}
  \caption{Non-Markovianity against number of collisions for $nn$ interactions, $nn$ and $2^{\text{nd}}$ $nn$ interactions, $nn$, $2^{\text{nd}}$ and $3^{\text{rd}}$ $nn$ interactions, and $nn$, $2^{\text{nd}}$, $3^{\text{rd}}$ and $4^{\text{th}}$ $nn$ interactions, with the interaction strengths $J_{ee}t=0.6(\pi/2)$, $J_{ee}t=0.43(\pi/2)$ and $J_{ee}t=0.33(\pi/2)$ respectively. The initial states that maximize $\mathcal{N}$ are $|\pm\rangle$ for all cases. \label{fig:collective}}
\end{figure}

In Fig.~\ref{fig:collective}, we present $\mathcal{N}$ versus the number of $s$-$e$ collisions. It is clear that extending the $e$-$e$ interactions and considering collective interactions significantly increases the degree of non-Markovianity as compared to just the $nn$ coupling. However, this increase is not monotonic in the number of interacting environmental qubits: the time evolution governed by $H_{ee}^{123}$ settles to a lower $\mathcal{N}$ value than that given by $H_{ee}^{12}$, but still greater than that obtained by $H_{ee}^1$. Including the $4^{\text{th}}$ $nn$ interaction further decreases $\mathcal{N}$, which however remains significantly higher than in the case of just $nn$ interaction. The mechanism behind this decrease may be the dilution of the system information that has leaked to the environment. At every $e$-$e$ interaction, some information about the original state of the first environmental qubit is transferred to the forthcoming subenvironments, together with some information from the system. Increasing the interaction length when such collective interactions are considered, results in increasing the number of times that an environmental qubit receives information from the system, before it interacts directly with the system qubit. As a result, in comparison with the separate interaction case, in the collective interaction case it is possible to reach higher values of non-Markovianity (cf. the y-axis ranges in Figs.~\ref{fig:separate} and \ref{fig:collective}). Furthermore, as the range of the $e$-$e$ interaction increases, the values at which $\mathcal{N}$ is saturated get closer and closer. One way to understand this result is that, as the number of collectively interacting subenvironments increases, we gradually approach the spin-boson model limit in the $e$-$e$ interactions, therefore the final values of $\mathcal{N}$ converge to the same number.

\subsection{Coherence}

Another important fact observed when considering collective $e$-$e$ interactions is that the number of collisions needed to reach the maximum amount of non-Markovianity increases almost by an order of magnitude, with respect to the $nn$ interaction case. The "time" it takes to reach the final configuration is also considerably higher as compared to the separate interaction case (cf. the x-axis ranges in Figs.~\ref{fig:separate} and Figs.~\ref{fig:collective}).

\begin{figure}[h]
\centering
  \includegraphics[width=.48\textwidth]{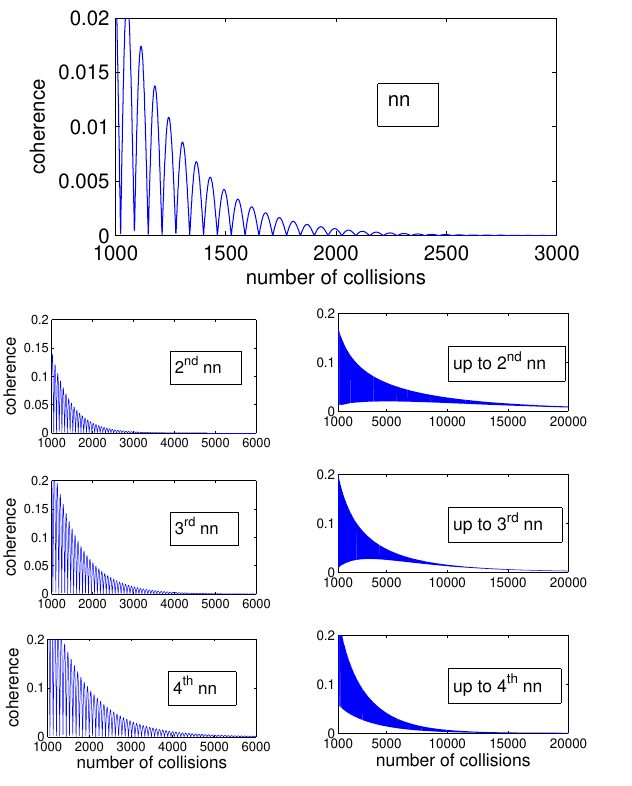}
  \caption{Plots of the $l_1$-norm of coherence against the number of collisions, for $nn$ (at the top), separate (left column) and collective (right column) interactions. From top to bottom the range of the $e$-$e$ interactions increases. We can see that the the number of $s$-$e$ collisions required for the coherence to vanish increases with longest living coherence being the one having the highest degree of non-Markovianity in both environment models. The frequency of oscillations in the collective interaction case is much higher than that of the separate interaction case. \label{fig:coherence}}
\end{figure}

The reason why the saturation of $\mathcal{N}$ happens later in the long-range separate and collective interaction cases, compared to the case with just $nn$ interaction, is that the revivals in the trace distance continue for a higher number of collisions, resulting in a more pronounced non-Markovianity of the dynamics. Since we interpret these revivals as the backflow of information from the environment to the system, we also looked for the effects of this information regain on the coherence contained in our system, which is initially in the fully coherent $|+\rangle$ or $|-\rangle$ state. Indeed, we observe a connection between the degree of non-Markovianity and the coherence, as shown in Fig.~\ref{fig:coherence}: prolonged oscillations in the trace distance, and therefore increased $\mathcal{N}$, are accompanied by prolonged oscillations in the coherence possessed by the system. We quantify the coherence with a recently introduced coherence measure, the $l_1$-norm of coherence
\begin{equation}\label{l1norm}
  C_{l_1}(\rho)=\sum_{i\neq j}|\rho_{i,j}|,
\end{equation}
which is nothing but the sum of the absolute values of the off-diagonal elements in the density matrix \cite{baumgratz}. $C_{l_1}$ satisfies all the criteria introduced in \cite{baumgratz} to be a valid measure of quantum coherence. Therefore, we have shown that the stronger the non-Markovianity in our system, the longer the time for which the coherence content will remain finite. In other words, by increasing the range of the interaction between subenvironments, we increase the number of collisions required for the complete decoherence of the system qubit, which is quite desirable in most practical cases. Comparing the separate interaction scenario with the collective one, we can conclude that in terms of coherence life-time the latter is much more advantageous than the former. Such a correlation between the non-Markovianity of a dynamical evolution and prolonged oscillations in the coherence have also been reported in a model constructed to understand the mechanism behind the long-lived coherence in photosynthetic complexes \cite{chen}. Furthermore, in Ref.~\cite{rivas} the interplay between coherence and non-Markovianity was examined in a refined spin-boson model, and it was shown that non-Markovianity causes revivals in the dynamics of coherence. A more detailed analysis in these lines of work can be found in \cite{streltsov}.


\subsection{System-environment correlations: mutual information}

We also investigated whether the trend of the non-Markovianity can be connected with the behaviour of the correlations created between the system and the subenvironment with whom it has just interacted. In order to quantify these correlations, we chose to look at the mutual information (MI) between $s$ and $e$ after they have interacted with each other:
\begin{equation}\label{mi}
  I(\rho_{se})=S(\rho_s)+S(\rho_e)-S(\rho_{se}),
\end{equation}
where $S(\rho)=-\text{tr}(\rho\text{ln}\rho)$ is the von Neumann entropy.

\begin{figure}[h]
\centering
  \includegraphics[width=.48\textwidth]{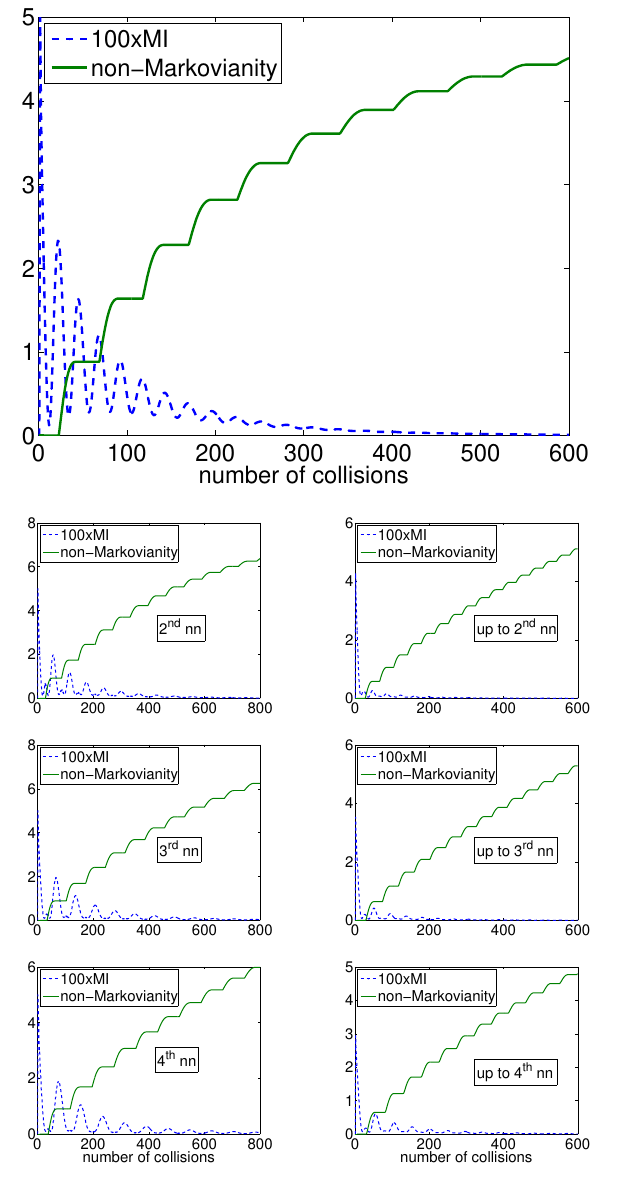}
  \caption{Mutual information and non-Markovianity against number of collisions for $nn$ (at the top), separate (left column) and collective (right column) interactions. From top to bottom the range of the $e$-$e$ interactions increases. \label{fig:MI}}
\end{figure}

We present our findings on the relation between the MI and $\mathcal{N}$ in Fig.~\ref{fig:MI}. First, one can immediately notice that the MI is significantly different from zero only around the first few hundred collisions, and approaches zero long before the non-Markovianity measure saturates to its final value. Furthermore, we can see that in the separate interaction case, as the interaction length is increased, the MI becomes more delocalized, and remains finite for a higher number of collisions. Such a behaviour of the MI seems correlated with the degree of non-Markovianity: the more the MI delocalizes (spreads) over the number of collisions, the higher is the increase between two plateaus of $\mathcal{N}$, which results directly in a higher degree of non-Markovianity. As the MI tends to zero, the increase of the non-Markovianity measure also slows down. Another correlation between the behaviours of MI and $\mathcal{N}$ is that the increases and plateaus of $\mathcal{N}$ occur in coincidence with the odd and even revivals of the MI respectively.

\subsection{System-environment correlations: a bound on the trace distance derivative}

Mutual information is not the only tool available to investigate the connection between quantum non-Markovianity and system-environment correlations. Ref.~\cite{Mazzola12} provides a link between the behaviour of the trace distance derivative $d D / dt$ and system-environment correlations, which are quantified using the matrix $\chi^{se}(t):=\rho^{se}(t)-\rho^{s}(t)\otimes\rho^{e}(t)$ as explained below; such matrix is identically zero when system and environment are completely uncorrelated. In Ref.~\cite{Mazzola12} the trace distance derivative is upper bounded by a quantity dependent explicitly on $\chi^{se}(t)$. The result is obtained in the weak coupling limit, under the assumptions of a system-environment interaction generated by the propagator $U_{t,t_0}=e^{-iH (t-t_0)/\hbar}$ for any initial time $t_0<t$, and of an initially uncorrelated system-environment state: let $\rho^s_{1,2}(t_0)$ be two arbitrary initial system states, and $\rho^e_{1}(t_0)=\rho^e_{2}(t_0)$ be identical initial environment states, than the joint initial states are $\rho^{se}_j(t_0) = \rho^{s}_j(t_0)\otimes\rho^{e}_j(t_0)$ for $j=1,2$.  These assumptions imply that the dynamics of the system is completely positive. Denoting the evolved marginal states of the system (environment) with $\rho^{s(e)}_j(t)=\text{Tr}_{e(s)}[U_{t,t_0}\rho_j^{se}(t_0)U_{t,t_0}^{\dag}]$, the bound on $d D / dt$ takes the form
\begin{equation}
\label{eq:mazzola1}
\begin{split}
&\frac{dD(t)}{dt} \le \,\frac{1}{2} \Big( B_{\text{Env}}(t) + B_{\text{Corr}}(t) \Big),
\\
&B_{\text{Env}}(t)=\big{\Vert} \tilde{\rho}^s_{1,1}(t) -  \tilde{\rho}^s_{1,2}(t) \big{\Vert},
\\
&B_{\text{Corr}}(t)= \big{\Vert} \tilde{\chi}^{s}_1(t) - \tilde{\chi}^{s}_2(t) \big{\Vert} ,
\end{split}
\end{equation}
where we employed the auxiliary states
\begin{equation}
\begin{split}
& \tilde{\rho}^s_{1,j}(t) = \text{Tr}_e \big{\{} \big[ H, \rho_1^{s}(t)\otimes \rho^e_j(t) \big] {\}} ,
\\
&\tilde{\chi}^{s}_j(t) = \text{Tr}_e \big{\{} \big[ H, \chi^{se}_j(t) \big] \big{\}}, \qquad \qquad j = 1,2,
\end{split}
\end{equation}
obtained from $\rho_1^{s}(t)\otimes \rho^e_j(t)$ and $\chi^{se}_j(t)$ by evolving them for an infinitesimally small time, through expansion of the operator $U$ to first order in $H$, and then taking the partial trace over the environment~\cite{footnote1}. In Eq.~(\ref{eq:mazzola1}), the term $B_{\text{Env}}(t)$ connects the emergence of non-Markovianity to the induced distinguishability in the environment states $\rho^e_1(t)$ and $\rho^e_2(t)$ which were initially identical. The term $B_{\text{Corr}}(t)$ instead accounts for the presence of system-environment correlations at time $t$, resulting from the previous evolution. Note that with our choice of initial conditions, at the beginning both $B_{\text{Env}}(t_0)$ and $B_{\text{Corr}}(t_0)$ are zero, and so $dD(t_0)/dt \le 0$.

\begin{figure}[h]
\centering
  \includegraphics[width=.48\textwidth]{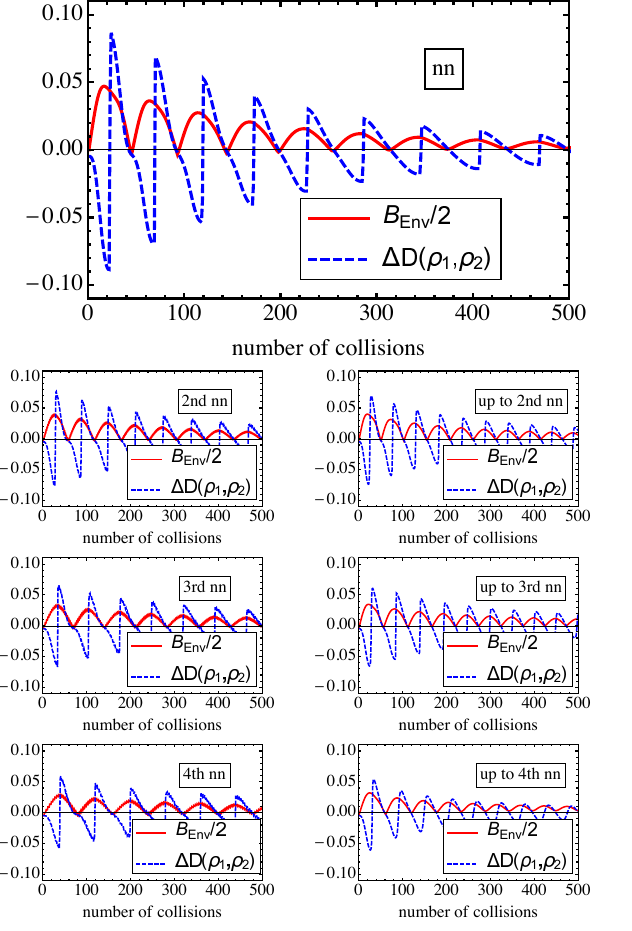}
  \caption{Upper bound on the trace distance derivative from Eq.~(\ref{eq:mazzola1}), for $nn$ (at the top), separate (left column) and collective (right column) interactions. From top to bottom the range of the $e$-$e$ interactions increases. Since only the $B_{\text{Env}}$ term gives a non-zero contribution, the bound clearly does not hold.\label{fig:SECbound}}
\end{figure}

The application of this bound to our dynamical model requires some care: in our implementation, after each iteration ($s$-$e$ followed by $e$-$e$ interaction), in order to prepare the states for the following step, system and environment are traced apart assigning to each of them the respective marginal state. This operation erases at every step all the $s$-$e$ correlations that may have been created, and causes the (discretized) term $B_{\text{Corr}}(k)$ to be identically 0 at each step $k$. Fig.~\ref{fig:SECbound} shows an example of the behaviour of the bound of Eq.~(\ref{eq:mazzola1}), to which only the term $B_{\text{Env}}(k)$ contributes. We plot it against the discretized trace distance derivative
\begin{equation}
\Delta D(\rho_1,\rho_2,k)= D(\rho^s_{1,k},\rho^s_{2,k})-D(\rho^s_{1,k-1},\rho^s_{2,k-1}).
\end{equation}

The term $B_{\text{Env}}$ alone is clearly not sufficient to bound the trace distance derivative, and this fact can be seen as a further proof of the relevance of system-environment correlations in non-Markovian quantum dynamics.

\begin{figure}[t]
\centering
  \includegraphics[width=.48\textwidth]{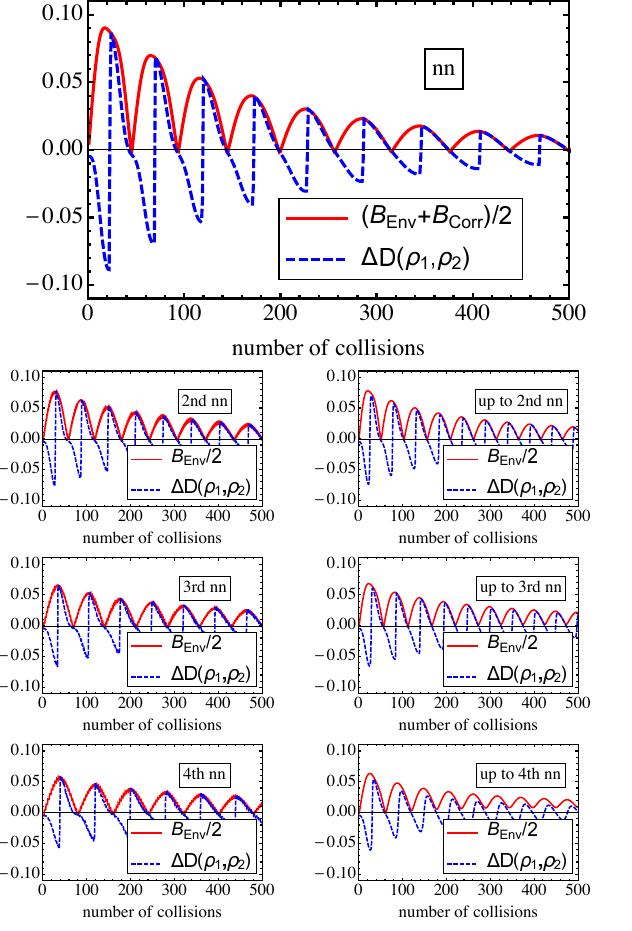}
  \caption{Upper bound on the trace distance derivative from Eq.~(\ref{eq:mazzola1}), computed now after each interaction and \textit{before} preparing the uncorrelated state for the following iteration. Displayed plots are for $nn$ (at the top), separate (left column) and collective (right column) interactions. From top to bottom the range of the $e$-$e$ interactions increases.  \label{fig:SECboundCorr}}
\end{figure}

In order to further investigate the situation, we implemented the computation of the bound in Eq.~(\ref{eq:mazzola1}) in a subtly different way. Instead of computing it at the \textit{beginning} of every step, now we compute it \textit{after} the $s$-$e$ and $e$-$e$ interactions occurred, and \textit{before} erasing the correlations, effectively making a fictitious evolution step as if we were able to carry over such correlations from one step to the following. The results are displayed in Fig.~\ref{fig:SECboundCorr} and now the bound is satisfied correctly~\cite{footnote2}. On a more fundamental level and in the spirit of~\cite{Mazzola12}, the bound computed straightforwardly with our discrete-time model, as shown in Fig.~\ref{fig:SECbound}, does not hold because of the following: the derivation of Eq.~(\ref{eq:mazzola1}) assumes that the evolved states $\rho_1^{s(e)}(k)$ and $\rho_2^{s(e)}(k)$ are connected to their respective initial states $\rho_1^{s(e)}(0)$ and $\rho_2^{s(e)}(0)$ uniquely by the unitary evolution given by $k$ subsequent applications of the 1-step unitaries $U_{se}$ and $U_{ee}$ from Eqs.~(\ref{eq:Uee}). However, in our model at every step, \textit{before} the application of $U_{se}$ and $U_{ee}$, the joint states $\rho_1^{se}(k)$ and $\rho_2^{se}(k)$ are each substituted with the tensor product of their two marginal system and environment states. Therefore, the overall process leading from $\rho_1^{s(e)}(0)$ and $\rho_2^{s(e)}(0)$ to $\rho_1^{s(e)}(k)$ and $\rho_2^{s(e)}(k)$ cannot be fully described by a unitary evolution, as required by the derivation of Eq.~(\ref{eq:mazzola1}). In the example of Fig.~\ref{fig:SECboundCorr} the dynamics is the same as in Fig.~\ref{fig:SECbound}, but the bound is computed at each step in such a way that this condition is satisfied for the current step.

\section{Conclusions}
\label{conclusions}

We have investigated the dynamics of a long-range collision model consisting of spin-$1/2$ particles. For practical applications and simplicity, we have modeled the $s$-$e$ and $e$-$e$ interactions as spin-spin interactions. We have considered two different models of $e$-$e$ interactions with varying interaction lengths. On the one hand, we have allowed the subenvironment which has interacted with the system qubit, to interact with its $nn$, $2^{\text{nd}}$ $nn$, $3^{\text{rd}}$ $nn$ or $4^{\text{th}}$ $nn$ separately. On the other hand, we have changed the $e$-$e$ to be a collective one, so that after interacting with the system, the subenvironment interacts with a collection of forthcoming environmental qubits, such as $nn$ $+2^{\text{nd}}$ $nn$, $nn$ $+2^{\text{nd}}$ $nn$ $+3^{\text{rd}}$ $nn$ or $nn$ $+2^{\text{nd}}$ $nn$ $+3^{\text{rd}}$ $nn$ $+4^{\text{th}}$ $nn$.

We have found that increasing the interactions beyond $nn$ immediately increases the non-Markovianity in the system dynamics. While in the separate interaction case this increase is linear with the distance between the interacting environmental qubits, in the collective interaction case it is non-monotonic with the number of qubits involved in the interaction increase. Moreover, it is possible to tune the degree of non-Markovianity by considering a scenario in which we consecutively apply separate $e$-$e$ interactions of different length and tunable strength. However, for collectively interacting environments, the interaction strength must be set to a very specific value in order to observe a non-Markovian dynamics. In both scenarios, the BLP measure of non-Markovianity saturates after a certain number of collisions between the system and the subenvironments. The number of collisions for which the saturation occurs is found to be related to the degree of non-Markovianity in the dynamics. We have seen that the higher the saturation value of non-Markovianity, the longer it takes to the system to reach that saturation value.

Another result that we have obtained is the observation of a direct connection between the degree non-Markovianity in the dynamics, and how fast the system loses coherence. By employing a recently proposed coherence measure, we have observed that the coherence of the system particle remains finite after a higher number of collisions in a dynamics that generates a higher degree of non-Markovianity. This may find application in relating resource theories of non-Markovianity and coherence, and allows for a non-Markovian route preserving coherence in a dynamical system.

Finally, we investigated the connection between non-Markovianity and system-environment correlations. The post collision mutual information shows how the odd and even peaks of revival of mutual information coincide respectively with the ramps and plateaus of the measure of non-Markovianity. Furthermore we computed an upper bound on the trace distance derivative, based on system-environment correlations and on the distinguishability induced in the environment. The necessity of both these contributions for the validity of the bound constitutes further evidence of the relevance of correlations in non-Markovian dynamics.

\begin{acknowledgments}
M. Pezzutto and M. Paternostro are grateful to Y. Omar for invaluable discussions. B\c{C} and \"{O}EM acknowledge support from a University Research Agreement between Lockheed-Martin Corp. and Koç University.
M. Pezzutto thanks the Centre for Theoretical Atomic, Molecular and Optical Physics, School of Mathematics and Physics, Queen's University Belfast for hospitality during the development and completion of this work, and also acknowledges the support from Funda\c{c}\~{a}o para a Ci\^{e}ncia e a Tecnologia (Portugal) and from the Doctoral Programme for the Physics and Mathematics of Information through scholarship SFRH/BD/52240/2013. M. Paternostro acknowledges financial support from the EU Collaborative Project TherMiQ (Grant Agreement 618074), the Julian Schwinger Foundation (grant number JSF-14-7-0000), and the DfE-SFI Investigator Programme (grant 15/IA/2864). B\c{C}, \"OEM and MP are supported by a Royal Society Newton Mobility Grant (grant NI160057). All authors gratefully acknowledge support from the COST Action MP1209 "Thermodynamics in the quantum regime".

\end{acknowledgments}


\begin{thebibliography}{99}
\bibitem{footnote1} One can derive an analogous bound using  $\rho_2^{s}(t)\otimes \rho^e_j(t)$ instead of  $\rho_1^{s}(t)\otimes \rho^e_j(t)$; in practice, we compute both and keep the smaller.
\bibitem{footnote2} Due to the difference in when the bound of Eq.~(14) is computed, the plots in Fig.~7, compared to those in Fig.~6, represent the bound computed at step $n+1$ instead of step $n$. To compensate for this, the plots of the trace distance derivative are shifted accordingly.
\end{thebibliography}
\end{document}